\documentclass{aa}  

\usepackage{graphicx}
\usepackage{subcaption}
\usepackage[colorlinks=true,     linkcolor=blue, citecolor=blue, filecolor=blue, urlcolor=blue]{hyperref}
\usepackage{txfonts}
\begin{document}

	\title{Long and short term variability of the possible nascent planetary nebula IRAS\,22568+6141: A late thermal pulse?}
	
	\author{Rold\'an A. Cala\inst{1}
        \and          
          Luis F. Miranda\inst{1}
         \and
         Jos\'e F. G\'omez\inst{1}
         \and
          Christophe Morisset\inst{2,3}
        \and
       Federico Soto\inst{2}
       \and
       Pedro F. Guill\'en\inst{4}
       \and 
       Roberto V\'azquez\inst{2}
          }

\institute{Instituto de Astrof\'{\i}sica de Andaluc\'{\i}a IAA--CSIC, Glorieta de la Astronom\'{\i}a s/n, 18008, Granada, Spain \\
\email{rcala@iaa.es} 
\and
Instituto de Astronom\'{\i}a, Universidad Nacional Aut\'onoma de M\'exico, Apdo. Postal 877, 22800 Ensenada, B.C., Mexico
\and
Instituto de Ciencias F\'{\i}sicas, Universidad Nacional Aut\'onoma de M\'exico, Av. Universidad s/n, 62210 Cuernavaca, Morelos, Mexico 
\and
Observatorio Astron\'omico Nacional, Instituto de Astronom\'{\i}a, Universidad Nacional Aut\'onoma de M\'exico, Apdo. Postal 106, 22800 Ensenada, B.C., Mexico 
         }
	
	\date{Received; accepted}
	
	
	\abstract{IRAS\,22568+6141 has been classified as a low-ionisation planetary nebula (PN) and presents non-thermal radio continuum emission, which could be a signature of nascent PNe. We present intermediate-resolution long-slit spectra obtained in 2021 and 2023, high-resolution long-slit spectra taken in 2023, and a light curve at the $r$ filter between 1953 and 2019, that reveal changes in IRAS\,22568+6141 with timescales of decades and a few years. The object underwent an energetic event around 1990 that suddenly increased its brightness which has been fading since then. A comparison with a published spectrum from 1988 shows an increase of the H$\beta$ flux in 2021 by factor of $\simeq$6 and the [O\,{\sc iii}] emission lines that were absent in 1988. Between 2021 and 2023 the H$\beta$ flux decreased by a factor of $\simeq$1.7, and the [O\,{\sc iii}] emission lines almost vanished. These results and the variability observed in other emission lines indicate that IRAS\,22568+6141 is recombining and cooling down between 2021 and 2023, and probably since 2005, as suggested by archival radio continuum and mid-IR observations. The intermediate- and high-resolution spectra show that the excitation of the emission lines is dominated by shocks in 2021 and 2023, and, probably, also in 1988, which may be related to the non-thermal radio continuum emission from the object. Although the variability might be due to changes in the physical conditions in the shocks or in a nova-like eruption, it accommodates better to that expected from a late thermal pulse, which is further suggested by a comparison with other similar objects. New observations and monitoring in the coming years are crucial to corroborate the origin of the variability. }

 
	\titlerunning{Variable nebula IRAS\,22568+6141: a \textit{late thermal pulse}?}
	
 \maketitle
	%
	
	\section{Introduction}
 
	Planetary nebulae (PNe) consist of circumstellar matter photoionised by hot central stars that evolve towards the cooling white dwarf phase. This ionised gas is the origin of most of the radiation from PNe. The emission at optical and infrared wavelengths is dominated by forbidden and recombination emission lines, while in the radio domain, the main emitting mechanism is bremsstrahlung (or free-free radiation) from free electrons in the plasma. This thermal continuum free-free radiation is characterised by a spectral index ($\alpha$, where S$_{\nu}$ $\propto$ $\nu$$^{\alpha}$) ranging from $+$2 (optically thick regime at low frequencies) to $-$0.1 \citep[optically thin regime at high frequencies; e.g.][]{aaq91} However, \cite{sua15} showed that the young PN IRAS 15103$-$5754 presents non-thermal radio continuum emission, with $\alpha$$\simeq$$-$0.54, compatible with synchrotron radiation. Several other possible PNe also present some evidence of non-thermal emission \citep{cer11,cer17}. The detection of non-thermal radio continuum emission in a PN could be an indication that it is at a nascent stage, since in more evolved PNe the free-free radiation would completely veil this non-thermal continuum emission.
 	
\begin{figure} [h]
		\centering
		\includegraphics[width=6.2cm]{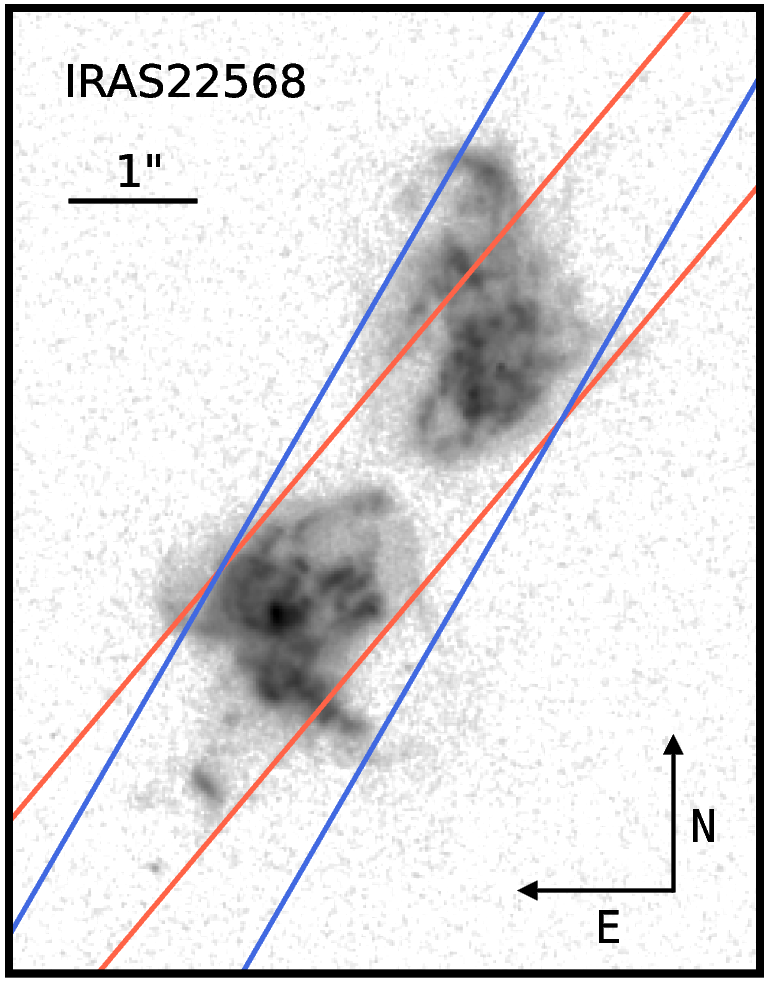}
		\caption{\textit{HST} (F606W) image of IRAS\,22568+6141 with the long-slit position used for our spectra represented in
			blue (width\,=\,2$''$, P.A. $-$30$\degr$), and the one used by GL+91 represented in red (width\,=\,1$\farcs$5,
			P.A. $-$40$\degr$).}
		\label{fig:i22568_hst_caha}
	\end{figure}

	IRAS\,22568+6141 (PN\,G110.1+01.9, $\alpha$(2000) =22h58m51.6s, $\delta$(2000)= +61$\degr$57$\arcmin$43.5$\arcsec$; hereafter IRAS22568) could be one of such nascent PNe. The optical spectrum presented by \cite{gar91} (hereafter GL+91) shows emission lines that are compatible with a low-excitation PN and is characterised by relatively prominent [N\,{\sc ii}] and absent [O\,{\sc iii}] emission lines. The reported radio continuum emission by \cite{cer11, cer17} suggests a negative spectral index ($\alpha$$\simeq$$-$0.35) that has been confirmed by G{\'o}mez et al. (in preparation) with quasi-simultaneous observations at different frequencies. Fig.\,\ref{fig:i22568_hst_caha} presents the Hubble Space Telescope ($HST$) image of IRAS22568 in the F606W filter. The nebula is bipolar and extends over $\simeq$\,8" along the main axis oriented at position angle P.A.$\simeq$\,$-$30$\degr$. The bipolar lobes are separated by a dark lane of width $\simeq$0$\farcs$25, where no optical emission is observed and seems to be obscuring the central star. Radio continuum observations by \cite{cer17} at 8.4 GHz resolve the two lobes and show emission from a central core between the lobes. 
	
    As part of a project to identify, confirm, and characterise new nascent PNe with non-thermal radio continuum emission, we obtained intermediate and high-resolution long-slit spectra of IRAS22568 at different epochs. In this paper we show that the source presents a peculiar variability in timescales of decades and a few years, as well as an internal kinematics not usually seen in PNe. The data presented here are crucial for understanding the nature of this possible nascent PN.
 
	  \begin{figure*}
		\centering
		\includegraphics[width=16.0cm]{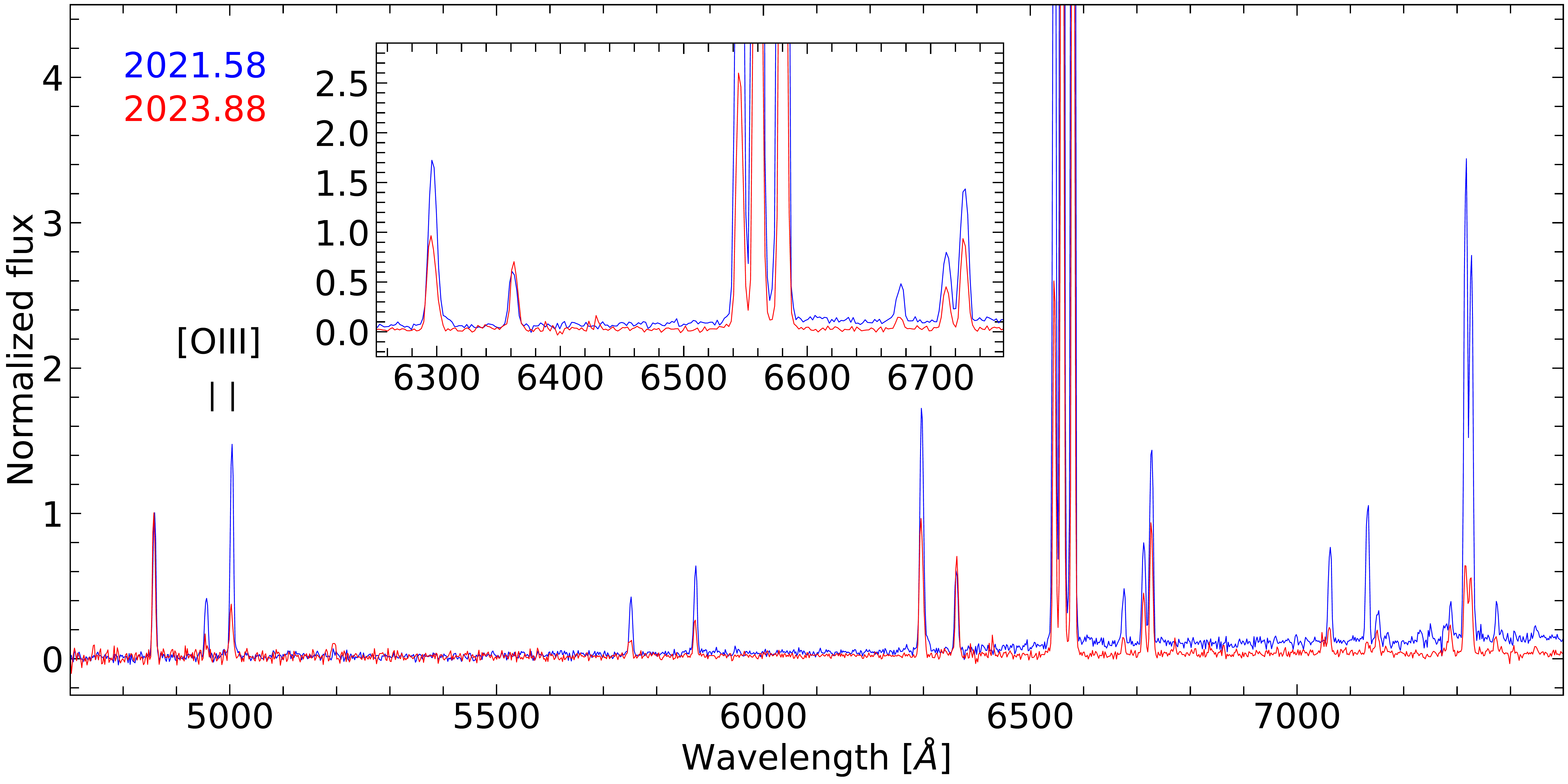}
        \caption{Intermediate-resolution optical spectra of IRAS\,22568+6141 obtained at the Calar Alto Observatory in 2021 (blue) and 2023 (red), normalized to peak flux of H$\beta$. The indicated [O\,{\sc iii}]$\lambda$4959,5007 emission lines were not detected in 1988 (see the text).}
		\label{fig:i22568_r_spec}
	\end{figure*}
	
	\section{Observations and data reduction}
	
	\subsection{Intermediate-resolution long-slit optical spectroscopy}
	\label{sec:obs_caha}
 
    We used the Calar Alto Faint Object Spectrograph (CAFOS) at the 2.2m telescope of the Calar Alto Observatory (Spain) on 2021 August 02 (proposal ID: F21-2.2-0.17; PI: R.\,A.\,Cala), and 2023 November 17 (proposal ID: DDT23B.320; PI: R.\,A.\,Cala) to obtain intermediate resolution long-slit spectra on IRAS22568. In Fig.\,\ref{fig:i22568_hst_caha} we show with blue lines the long-slit (width\,=\,2$\farcs$0) oriented at P.A. $-$30$\degr$ along the major nebular axis. In both epochs the sky was photometric during the observations with a mean seeing of 2$\farcs$1, and in both epochs we observed the standard star BD+28$\degr$4211 for flux calibration. The spectra were recorded in a SITe CCD with 2048$\times$2048\,pixel$^2$. The spatial scale in the detector is 0.53\,arcsec\,pixel$^{-1}$. In 2021/2023 we obtained four/three spectra using grism B-100 (3200-6200\,$\AA$) with 1800\,s exposure time each, and combined them into a single spectrum for each epoch. Also, in each epoch we obtained one spectrum with grism R-100 (5800-9600\,$\AA$) and 1800\,s of exposure time. The spectra were calibrated following standard procedures in {\sc iraf}. We obtained a dispersion of $\sim$2\,$\AA$\,pixel$^{-1}$ after the wavelength calibration. PyNeb \citep[version 1.1.19;][]{lur15} was used to obtain the physical properties of the ionised gas, and Table\,\ref{tab:atomic_data} lists the employed atomic constants. 

	\subsection{High-resolution long-slit optical spectroscopy}

    High-resolution long-slit spectroscopy of IRAS 22568 was conducted on 2023 August 5--7 using the Manchester Echelle Spectrometer (MES) at the 2.1\,m telescope on the San Pedro Mártir Observatory (Mexico) (proposal ID: FM-15; PI: R.\,V{\'a}zquez). The long-slit (2" width) was oriented at P.A. $-$30$\degr$ (Fig.\,\ref{fig:i22568_hst_caha}). The spectra were recorded in an E2V-4240 CCD with 1024$\times$1024\,pixel$^2$ and 4$\times$4 binning, resulting in a spatial scale of 0$\farcs$704\,pixel$^{-1}$. Spectra of the H$\alpha$, [N\,{\sc ii}]$\lambda$6584, [S\,{\sc ii}]$\lambda$6730, and [O\,{\sc iii}]$\lambda$5007 emission lines (hereafter [N\,{\sc ii}], [S\,{\sc ii}], and [O\,{\sc iii}]) were obtained with exposure times of 1800, 1800, 3600, and 3600\,s, respectively. A Th-Ar lamp spectrum was obtained immediately after each source spectrum for wavelength calibration. The spectra were reduced with {\sc iraf}. The average seeing during the observations was $\sim$1$\farcs$8, and the spectral resolution is $\sim$11\,km\,s$^{-1}$, as indicated by the full width at half maximum (FWHM) of the emission lines in the ThAr spectra.

	\section{Results}
 
	\subsection{Spectral variability and physical properties of the ionised gas}
    \label{section:physical}

    Fig.\,\ref{fig:i22568_r_spec} shows the intermediate-resolution spectra of IRAS22568 integrated over its two bipolar lobes. A simple look to our Fig.\,2 and Fig.\,3 in GL+91 reveals significant differences between the 1988, 2021, and 2023 spectra. Although our spectra resolve the two bipolar lobes of the object, for a better comparison between the spectra at the three epochs, we used the observed emission line fluxes in the integrated spectra for the analysis. The observed fluxes in 1988 can be found in GL+91 (their Table 2, column 3), and those in 2021 and 2023 are listed in Table\,\ref{tab:lines_uncorrected} . 
    
    We used the interstellar extinction law of \cite{sea79} to obtain the logarithmic extinction coefficient at H$\beta$, $c$(H$\beta$), and hereby, the intrinsic emission line intensities in the three epochs. The $c$(H$\beta$) was derived from the H$\alpha$ and H$\beta$ emission lines and a theoretical H$\alpha$/H$\beta$ line intensity ratio of 2.85. We noticed that a similar $c$(H$\beta$) is obtained from the Paschen\,9 (P\,9) and H$\beta$ emission lines, and a theoretical P\,9/H$\beta$ line intensity ratio of 18.4$\times$10$^{-3}$. The intrinsic line intensities, H$\beta$ flux, and $c$(H$\beta$) at the three epochs are listed in Table\,\ref{table:em_lines}. The differences between the intrinsic line intensities in 1988 obtained by us (listed in our Table\,\ref{table:em_lines}, column\,4) and those obtained by GL+91 (listed in their Table\,2, column\,4) are not significant and due to the different extinction law employed.

 \begin{table} [h]
		\caption{Intrinsic line intensities ($I$(H$\beta$)=100) in IRAS22568.}             
		\label{table:em_lines}      
		\centering                          
		\begin{tabular}{l c c c c}        
			\hline\hline                 
			Emission & I($\lambda$) & I($\lambda$) & I($\lambda$)     \\
			line &  1988.58 & 2021.58 & 2023.88   \\ 
			\hline                        
			4101\,H$\delta$	&	--\tablefootmark{a}	&	40.71$\pm$2.16	&	--\tablefootmark{a}	\\
            4340\,H$\gamma$	&	--\tablefootmark{a} 	&	57.12$\pm$3.72	&	--\tablefootmark{a}	\\
            4363\,[O\,{\sc iii}]	&	--\tablefootmark{a} 	&	8.81$\pm$2.29	&	--\tablefootmark{a} \\
            4471\,He\,{\sc i}	&	--\tablefootmark{a} 	&	7.6$\pm$0.7	&	--\tablefootmark{a}	\\
            4861\,H$\beta$	&	100$\pm$10 	&	100.0$\pm$1.6	&	100.0$\pm$2.5	\\
            4959\,[O\,{\sc iii}]	&	--\tablefootmark{a} 	&	40.70$\pm$1.15	&	15.02$\pm$1.60	\\
            5007\,[O\,{\sc iii}]	&	--\tablefootmark{a} 	&	119.09$\pm$1.35	&	40.58$\pm$2.01	\\
            5755\,[N\,{\sc ii}]	&	8.05$\pm$0.79 	&	11.00$\pm$0.28	&	7.65$\pm$0.30	\\
            5876\,He\,{\sc i}	&	11.73$\pm$1.11 	&	13.83$\pm$0.25	&	12.86$\pm$0.40	\\
            6300\,[O\,{\sc i}]	&	25.89$\pm$1.32 	&	22.25$\pm$0.19	&	67.16$\pm$0.68	\\
            6363\,[O\,{\sc i}]	&	10.03$\pm$0.49 	&	8.10$\pm$0.18	&	26.88$\pm$0.64	\\
            6548\,[N\,{\sc ii}]	&	73.9$\pm$1.3 	&	70.04$\pm$2.46	&	80.33$\pm$1.67	\\
            6563\,H$\alpha$	&	285.0$\pm$5.1 	&	285.0$\pm$2.5	&	285.0$\pm$1.6	\\
            6584\,[N\,{\sc ii}]	&	215.7$\pm$4.7	&	220.0$\pm$2.4	&	235.0$\pm$1.6	\\
            6678\,He\,{\sc i}	&	2.80$\pm$0.39 	&	3.69$\pm$0.11	&	4.28$\pm$0.26	\\
            6716\,[S\,{\sc ii}]	&	8.86$\pm$0.81 	&	7.01$\pm$0.12	&	13.07$\pm$0.27	\\
            6731\,[S\,{\sc ii}]	&	17.46$\pm$1.32 	&	13.44$\pm$0.13	&	26.00$\pm$0.28	\\
            7002\,O\,{\sc i}	&	-- \tablefootmark{a} 	&	0.49$\pm$0.02 &	--\tablefootmark{a}	\\
            7065\,He\,{\sc i}	&	4.2$\pm$0.4 	&	4.63$\pm$0.08	&	4.69$\pm$0.19	\\
            7137\,[Ar\,{\sc iii}]	&	6.1$\pm$0.9 	&	6.37$\pm$0.08	&	3.19$\pm$0.19	\\
            7150\,[Fe\,{\sc ii}]	&	--\tablefootmark{a} 	&	1.49$\pm$0.07	&	4.20$\pm$0.19	\\
            7230\,[Fe\,{\sc ii}]	&	--\tablefootmark{a} 	&	0.59$\pm$0.04	&	3.53$\pm$0.26	\\
            7254\,O\,{\sc i}	&	--\tablefootmark{a} 	&	0.54$\pm$0.04	&	3.24$\pm$0.25	\\
            7281\,He\,{\sc i}	&	--\tablefootmark{a} 	&	0.75$\pm$0.09	&	5.90$\pm$0.21	\\
            7319\,[O\,{\sc ii}]	&	blended\tablefootmark{b} 	&	19.98$\pm$0.15	&	15.16$\pm$0.22	\\
            7330\,[O\,{\sc ii}]	&	blended\tablefootmark{b} 	&	15.37$\pm$0.14	&	13.95$\pm$0.23	\\
            7378\,[Ni\,{\sc ii}]	&	1.73$\pm$0.43 	&	1.31$\pm$0.14	&	3.05$\pm$0.15	\\
            7447\,[Fe\,{\sc ii}]	&	--\tablefootmark{a} 	&	0.55$\pm$0.030	&	1.79$\pm$0.18	\\
            7751\,[Ar\,{\sc iii}]	&	--\tablefootmark{a} 	&	1.83$\pm$0.07	&	0.67$\pm$0.04	\\
            8446\,O\,{\sc i}	&	blended\tablefootmark{c} 	&	3.33$\pm$0.06	&	1.61$\pm$0.11	\\
            8467\,P\,17	&	blended\tablefootmark{c} 	&	0.81$\pm$0.05	&	0.71$\pm$0.04	\\
            8502\,P\,16	&	--\tablefootmark{a} 	&	0.85$\pm$0.06	&	1.54$\pm$0.11	\\
            8545\,P\,15	&	--\tablefootmark{a} 	&	1.09$\pm$0.06	&	1.97$\pm$0.14	\\
            8578\,[Cl\,{\sc ii}]	&	--\tablefootmark{a} 	&	0.77$\pm$0.06	&	2.26$\pm$0.12	\\
            8598\,P\,14	&	--\tablefootmark{a} 	&	0.79$\pm$0.06	&	1.76$\pm$0.15	\\
            8605\,[Fe\,{\sc ii}]	&	--\tablefootmark{a} 	&	1.31$\pm$0.06	&	4.37$\pm$0.12	\\
            8665\,P\,13	&	--\tablefootmark{a} 	&	1.27$\pm$0.06	&	2.26$\pm$0.14	\\
            8750\,P\,12	&	--\tablefootmark{a} 	&	0.96$\pm$0.05	&	2.29$\pm$0.13	\\
            8863\,P\,11	&	--\tablefootmark{a} 	&	1.53$\pm$0.05	&	2.71$\pm$0.11	\\
            8890 [Fe\,{\sc ii}]	&	--\tablefootmark{a} 	&	0.91$\pm$0.05	&	2.29$\pm$0.11	\\
            9015\,P\,10	&	1.57$\pm$0.24 	&	1.83$\pm$0.07	&	1.63$\pm$0.07	\\
            9069\,[S\,{\sc iii}]	&	9.41$\pm$0.86 	&	14.42$\pm$0.07	&	5.15$\pm$0.09	\\
            9229\,P\,9	&	3.45$\pm$0.43 	&	3.44$\pm$0.08	&	5.10$\pm$0.15	\\
            9535\,[S\,{\sc iii}]	&	22.34$\pm$1.52\tablefootmark{d} &	32.22$\pm$0.19	&	11.86$\pm$0.30	\\
			\hline 
			$F$(H$\beta$)\tablefootmark{e} & 1.16$\pm$0.12 & 6.88$\pm$0.11 &  4.11$\pm$0.11\\
			  
			$c$(H$\beta$)  &  3.70$\pm$0.16 & 3.05$\pm$0.05 & 1.66$\pm$0.04 \\
			Av\tablefootmark{f}\,(mag) & 7.88$\pm$0.34 & 6.48$\pm$0.10 & 3.53$\pm$0.09 \\
   
			\hline                                   
		\end{tabular}
		\tablefoot{
            \tablefootmark{a}{Not detected by GL+91 or in 2023.95.}
  			\tablefootmark{b}{[O\,{\sc ii}]\,$\lambda$7319,7330 are blended, with $I$(sum)=32.63$\pm$2.26.}
			\tablefootmark{c}{O\,{\sc i}\,$\lambda$8446 and P\,17\,$\lambda8467$ are blended with $I$(sum)=6.09$\pm$0.61.}
            \tablefootmark{d}{Emission line blended with P\,8\,$\lambda9546$.}
            \tablefootmark{e}{in units of 10$^{-15}$ erg\,cm$^{-2}$\,s$^{-1}$.}
			\tablefootmark{f}{Nebular extinction, see the text for details.}	
		}
	\end{table}
 
    \begin{table}
		\caption{Electron temperature and electron density of the ionised gas in IRAS 22568+6141 in the three epochs.} 
		\label{table:physical}      
		\centering                          
		\begin{tabular}{l c c c}        
			
			\hline\hline
			& 1988.58\tablefootmark{a} & 2021.58  & 2023.88 \\               
			\hline
            $T_{\rm e}$ [N\,{\sc ii}]\,(K) & 13600$\pm$1000 & 17300$\pm$600  & 12400$\pm$300 \\
			$T_{\rm e}$ [O\,{\sc iii}]\,(K)  & --\tablefootmark{b} & 38400\,$^{+17000} _{-10500}$ & --\tablefootmark{c} \\
			$N_{\rm e}$ [S\,{\sc ii}]\,(cm$^{-3}$)  & 12100\,$^{+} _{-8500}$ & 10600$\pm$2100 & 12900\,$^{+4500} _{-2900}$ \\
			\hline                                   
		\end{tabular}
		\tablefoot{
			\tablefootmark{a}{No reliable upper limit for the electron density could be obtained.}
            \tablefootmark{b}{[O\,{\sc iii}] emission lines not detected.}
            \tablefootmark{c}{[O\,{\sc iii}]$\lambda$4363 not detected.}
			}
		
	\end{table}
 
    The most astonishing result is the presence in 2021 of the [O\,{\sc iii}]\,$\lambda$4363,4959,5007 emission lines in the nebula with $I$([O\,{\sc iii}])/$I$(H$\beta$)$\simeq$1.7, which were absent in 1988 (GL+91). Using the intensity of the weakest emission line detected in the spectrum from 1988 ($I$(P\,10)/$I$(H$\beta$)$\simeq$0.016; Table\,\ref{table:em_lines}) as an upper limit to the detection of the [O\,{\sc iii}] lines, they have emerged between 1988 and 2021 or, at least, their intensity has increased by more than two orders of magnitude in that time span. Further impressive is the subsequent variability of the [O\,{\sc iii}]\,$\lambda$4959,5007 emission lines whose intensity relative to H$\beta$ has decreased by a factor of $\simeq$3 between 2021 and 2023, while the [O\,{\sc iii}]$\lambda$4363 line is not detected in the last epoch.

    More emission properties of the ionised gas have also substantially changed in IRAS22568. Table\,\ref{table:em_lines} shows that the H$\beta$ flux increased by a factor $\simeq$6 between 1988 and 2021. This value should be taken with some care considering that the long-slits used in 1988 and 2021 do not cover exactly the same regions of the object (Fig.\,\ref{fig:i22568_hst_caha}), as well as the different extinction (see below) and sensitivity of the observations. However, a large increase of the H$\beta$ flux between 1988 and 2021 should be considered real. Between 2021 and 2023 the H$\beta$ flux decreased by a factor of $\simeq$1.7 and we note that these two spectra were obtained under almost identical conditions (Sect.\,\ref{sec:obs_caha}).
  
    The value of $c$(H$\beta$) (Table\,\ref{table:em_lines}) is decreasing since 1988 by factors of $\simeq$1.06 between 1988 and 2021, and $\simeq$1.8 between 2021 and 2023. Given that the nebular reddening E\,(B--V) and the nebular extinction Av are related by $c$(H$\beta$)=1.45$\times$E\,(B--V), and $c$(H$\beta$)=0.47$\times$Av \citep[using the][extinction law]{sea79}, we obtain Av(1988)=7.88$\pm$0.34\,mag, Av(2021)=6.48$\pm$0.10\,mag, and Av(2023)=3.53$\pm$0.10\,mag. Therefore, the nebular extinction of IRAS22568 decreased 1.40$\pm$0.35\,mag between 1988 and 2021, and 2.95$\pm$0.10\,mag between 2021 and 2023. 
        
    Other emission lines present variability of their intensity relative to H$\beta$ between 1988 and 2023. In the cases of very weak emission lines (e.g., H$\gamma$, H$\delta$, [Cl\,{\sc ii}], and [Fe\,{\sc ii}]) that were not present in the 1988 spectrum, their detection in 2021 and/or 2023 may be due to a combination of increased nebular flux and better sensitivity in our spectra. In the cases of the [S\,{\sc ii}], [O\,{\sc i}], [O\,{\sc ii}], and He\,{\sc i} emission lines the variability can be considered real because these lines are relatively strong and detected in all three epochs. Moreover, an inspection of the 2021 and 2023 line intensities reveals that IRAS22568 has recombined in that time span. This is directly recognizable in (a) the decrease of the O$^{2+}$, O$^{+}$, and S$^{2+}$ line intensities (relative to H$\beta$) in favour of those of O$^{+}$, O$^{0}$, and S$^{+}$, respectively, (b) the decreasing intensity of the Ar$^{2+}$ emission lines, (c) the increasing intensity of the Fe$^{+}$ and Ni$^{+}$ ones, and (d) the decrease of the Balmer emission intensities. The [N\,{\sc ii}]\,$\lambda$6548,6583 emission lines present a slight increase in their relative intensities between 2021 and 2023 probably due to recombination from N$^{2+}$ that proceeds faster than recombination from N$^{+}$ to N$^{0}$. Contrarily to that observed in the nebular [N\,{\sc ii}] lines, the auroral [N\,{\sc ii}]\,$\lambda$5755 line intensity clearly decreases between 2021 and 2023, implying that the nebula is also cooling down.
 
    To calculate the electron temperature ($T_{\rm e}$) we use the [N\,{\sc ii}](6584+6548)/5755 and [O\,{\sc ii}](4959+5007)/4363 emission lines ratios, while the electron density ($N_{\rm e}$) was obtained from the [S\,{\sc ii}](6716/6731) line intensity ratio. These values were obtained making use of {\sc crosstemden} in PyNeb. The results are listed in Table\,\ref{table:physical}. The values of $T_{\rm e}$([N\,{\sc ii}]) are consistent with an increase of the excitation between 1988 and 2021, and with cooling between 2021 and 2023. In 2021, when the [O\,{\sc ii}]\,$\lambda$4363 is detected, we obtained $T_{\rm e}$([O\,{\sc iii}])\,$\simeq$\,38400\,K and $T_{\rm e}$([N\,{\sc ii}])\,$\simeq$\,17300\,K. $T_{\rm e}$([O\,{\sc iii}]) is anomalously high and very much higher than $T_{\rm e}$([N\,{\sc ii}]). Such values can hardly be explained with photoionisation but clearly point out to shock excitation. This will be discussed in detail in Sect.\,\ref{sec:bow_shocks}. In 2023 we obtained an upper limit for the [O\,{\sc iii}]\,$\lambda$4363 line intensity (Table\,\ref{tab:lines_uncorrected}), which does not allow us to derive $T_{\rm e}$([O\,{\sc iii}]), while the [O\,{\sc iii}] lines were not detected in 1988. 

    As of the electron density (Table\,\ref{table:physical}), the uncertainties in the line intensities in 1988 do not allow us to obtain an upper limit of $N_{\rm e}$, suggesting that the nebula could have been in a high-density regime (see GL+91). On the other hand, the nominal values of $N_{\rm e}$ agree with each other within the errors. We noted that the values of $N_{\rm e}$ are averaged values integrated along the line of sight. The knotty structure of IRAS2258 suggests that $N_{\rm e}$ may present a relatively large range of values throughout the nebula \citep[see e.g.][]{lee22, hyu23}.
    
    With the values obtained for $N_{\rm e}$ (Table\,\ref{table:physical}) we can estimate the recombination timescales assuming a recombination rate per ion $R_{\rm i}\simeq$ 10$^{5}$\,yr\,/\,(Z$^{2}$\,$N_{\rm e}$), where Z is the ionic charge \citep{ost89}. The recombination timescales of single and double ionised species result to be 8--12 and 2--3\,yr, respectively, which are consistent with the fact that the [O\,{\sc iii}] emission lines have almost vanished between 2021 and 2023, while the Balmer ones show a more moderate decrease in intensity between those two epochs.

\subsection{Light curve}\label{sec_photo}

In order to further investigate the optical variability of IRAS22568, we have searched for available photometric broadband \textit{r}-filter observations in the literature and public surveys. We found six photometric values for the \textit{r} magnitude from 1953 to 2020, that are reported in Table\,\ref{table:photo}. However, the characteristics of the filters are not the same in the six epochs. The PanSTARRS filter is similar to that of GL+91, thus we transformed the magnitudes of SDSS to the PanSTARRS system magnitude \citep[see][]{pan19}. The filter of ZTF is broader and has a transmission higher than the other filters, thus the magnitude from 2019.5 should be seen as a lower limit to the \textit{r} magnitude measured in PanSTARRS. Finally, the POSS-I and POSS-II filters are narrower than the other filters, and their \textit{r} magnitudes should be considered as upper limits to the \textit{r} magnitude obtained in PanSTARRS.

\begin{table}
        \setlength{\tabcolsep}{3.8pt}
		\caption{Multiepoch broadband \textit{r}-filter photometry on IRAS22568.} 
		\label{table:photo}    
		\centering                          
		\begin{tabular}{l c c c}        
			
			\hline\hline
			\rule{0pt}{2.3ex}
			 Epoch  & Survey \tablefootmark{a}	  & \textit{r} (mag) & Reference \\               
			\hline
			1953.8 & POSS-I & 16.16$\pm$0.25  & \cite{mon03} \\
			1990.5 & -- & 16.7$\pm$0.3  & GL+91 \\
			1991.7 & POSS-II & 15.15$\pm$0.25  & \cite{mon03} \\
			2003.7 & SDSS & 15.92$\pm$0.02  & \cite{sdss09} \\
			2012.7 & PanSTARRS  & 16.40$\pm$0.01 & \cite{pan19}\\
			2019.5 & ZTF & 16.15$\pm$0.03  & \cite{ztf18} \\
			\hline                                   
		\end{tabular}
		\tablefoot{
			\tablefootmark{a}{Palomar Observatory Sky Survey first (POSS-I) and second (POSS-II) release, Sloan Digital Sky Survey (SDSS), PanSTARRS survey, and Zwicky Transient Facility (ZTF) survey.}}
		
	\end{table}

\begin{figure}
		\centering
        \includegraphics[width=8.0cm]{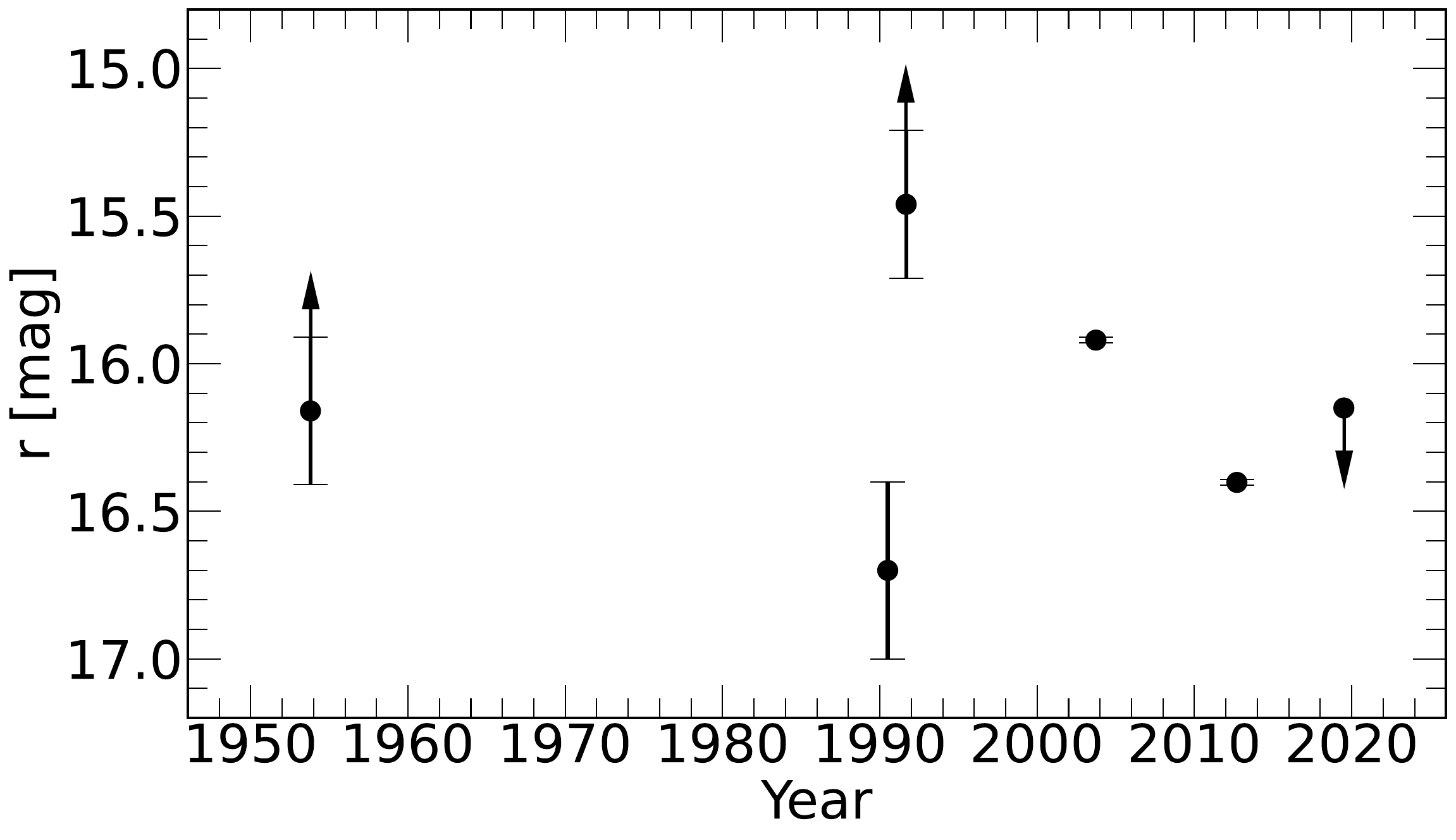}
		\caption{Light curve of IRAS22568 in the \textit{r}-filter. The arrows represent lower or upper limits. See the text for details.}
		\label{fig:i22568_r_photo}
	\end{figure}

Fig.\,\ref{fig:i22568_r_photo} shows the \textit{r} light curve of IRAS22568. Despite the limited temporal sampling and the lower/upper limits at some epochs, some trends can be seen. The \textit{r} magnitude seems to have decreased $\geq$0.54\,mag between 1953 and 1990.5, suddenly rose $\geq$1.5\,mag between 1990.5 and 1991.7, and presented a steady decreasing tendency of $\geq$1\,mag between 1991.7 and 2019.5. Our 2021 and 2023 spectra are consistent with this decrease (Table\,\ref{table:em_lines}). Moreover, the flux density of the radio continuum emission has also been fading between 2005 and 2012 \citep{cer17}.   

\subsection{Internal kinematics}\label{sec_internal}

\begin{figure}
		\centering
		\begin{subfigure}{0.89\linewidth}
			\includegraphics[width=\linewidth]{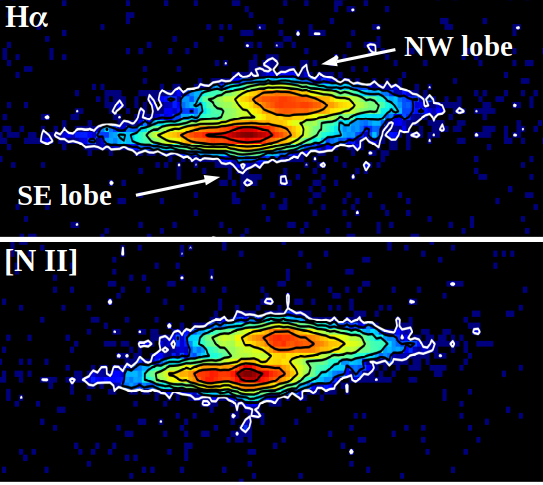}
		\end{subfigure}
		\begin{subfigure}{0.89\linewidth}
			\includegraphics[width=\linewidth]{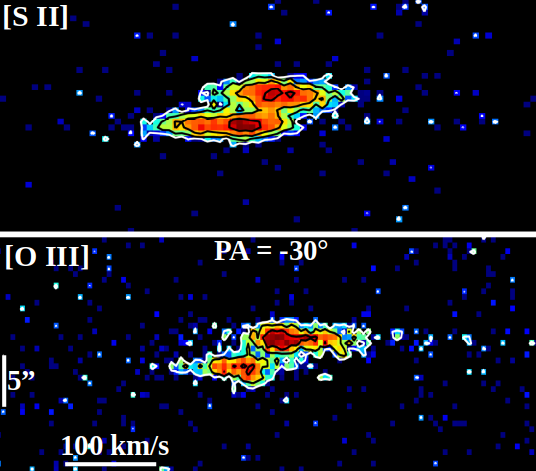}
		\end{subfigure}
		\caption{Position-velocity maps of the H$\alpha$, [N\,{\sc ii}], [S\,{\sc ii}], and [O\,{\sc iii}] emission lines obtained at a position angle P.A.\,$-$30 in 2023.68. The increment step of the contours is 3$\sigma$\,$\times$\,2$^{n}$, starting from $n$\,=\,0 (white colour). The south-eastern lobe (SE lobe) and north-western lobe (NW lobe) are indicated.}
		\label{fig:pv_map}
	\end{figure}

\begin{figure}
		\centering
		\begin{subfigure}{0.9\linewidth}
		\includegraphics[angle=0, width=\linewidth]{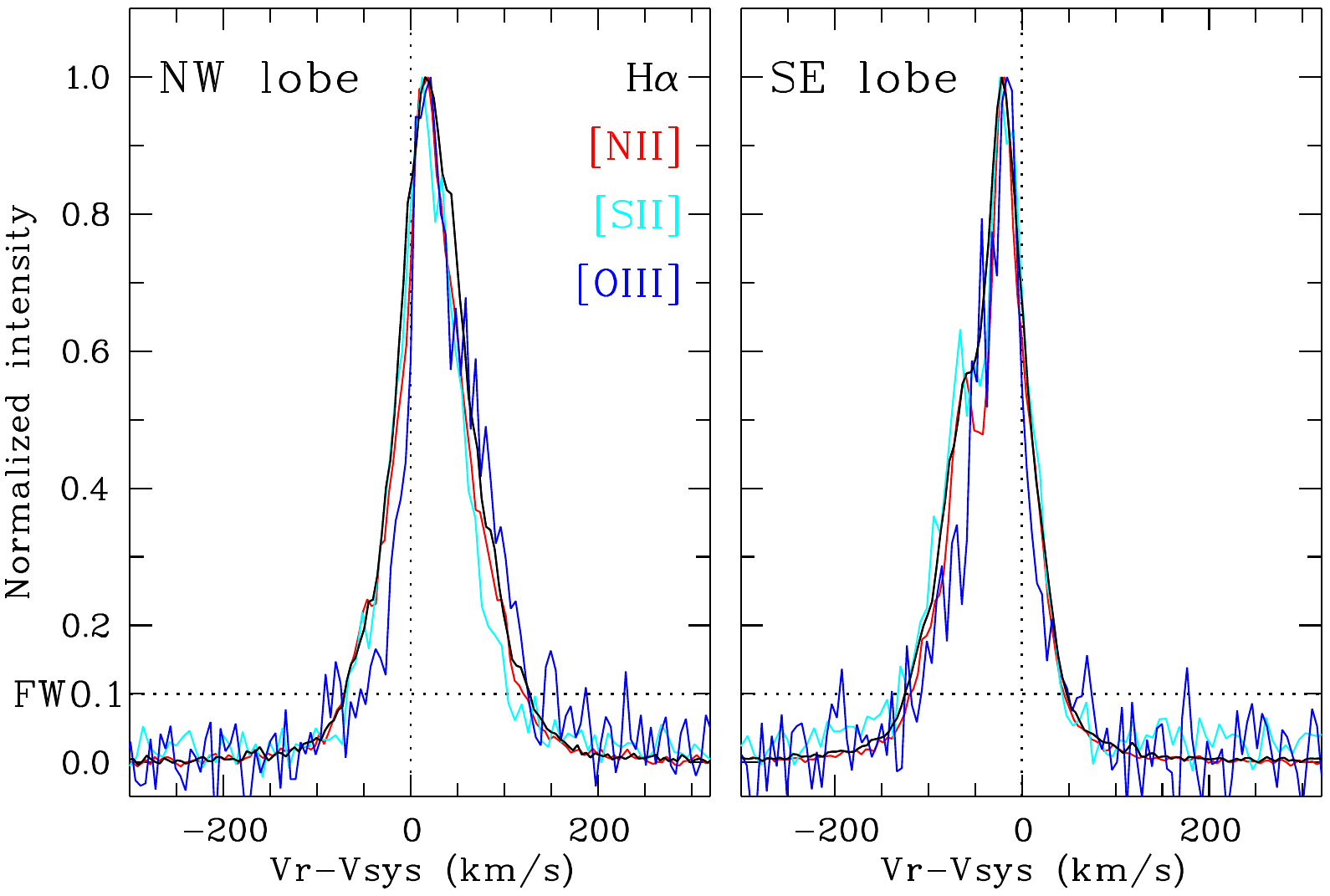}
		\end{subfigure}
		\caption{Normalized line profiles of the H$\alpha$, [N\,{\sc ii}], [S\,{\sc ii}], and [O\,{\sc iii}] emission lines of the north-western (NW) and south-eastern (SE) lobes. The full-width at 10\% the peak intensity (FW0.1) is shown as an horizontal dotted line.}
		\label{fig:line_profiles}
	\end{figure}

Fig.\,\ref{fig:pv_map} displays position-velocity (PV) maps of the H$\alpha$, [N\,{\sc ii}], [S\,{\sc ii}], and [O\,{\sc iii}] emission lines derived from the high-resolution long-slit MES spectra obtained in 2023.68, which are close in time to the intermediate resolution spectra from 2023.88. The MES spectra spatially resolve the NW and SE lobes of IRAS22568. The emission lines arise in two relatively compact regions and are characterised by a large velocity width that reach up to 450 and 380 km\,s$^{-1}$ in H$\alpha$ and [N\,{\sc ii}], respectively, as measured at the 3$\sigma$ level above the background (white contours in Figure\,\ref{fig:pv_map}). The [S\,{\sc ii}] and [O\,{\sc iii}] emissions are much fainter than the H$\alpha$ and [N\,{\sc ii}] ones, although they seem to present a similar profile. Fig.\,\ref{fig:line_profiles} presents the emission line profiles as obtained by integrating the emission of each lobe in a region of $\simeq$2" centred on the emission peaks. Despite the differences in S/N, the [S\,{\sc ii}] and [O\,{\sc iii}] emission lines also exhibit large velocity widths that seem to be comparable to those observed in H$\alpha$ and [N\,{\sc ii}]. 

\begin{table*}
		\caption{Properties of the H$\alpha$, [N\,{\sc ii}], [S\,{\sc ii}], and [O\,{\sc iii}] emission lines in the high resolution spectra.} 
		\label{table:bow_shocks}      
		\centering                          
		\begin{tabular}{l c c c c c c c c c c}        
			\hline\hline
			\rule{0pt}{2.5ex}                 
			& \multicolumn{3}{c}{NW lobe\tablefootmark{a}} &  \multicolumn{3}{c}{SE lobe\tablefootmark{a}}  \\
            \hline
			\rule{0pt}{2.5ex}                 
			& $V_{\rm peak}$ \tablefootmark{b} & $V_{\rm min}$ \tablefootmark{c} & $V_{\rm max}$ \tablefootmark{c} & $V_{\rm peak}$ \tablefootmark{b} & $V_{\rm min}$ \tablefootmark{c} & $V_{\rm max}$ \tablefootmark{c} & $\delta R$ \tablefootmark{d} & $V_{\rm sys}$ \tablefootmark{e} & $I_{\rm NW}/I_{\rm SE}$\tablefootmark{f} \\
			Emission line &		km\,s$^{-1}$		&  km\,s$^{-1}$ &	 km\,s$^{-1}$	&  km\,s$^{-1}$  & km\,s$^{-1}$ & km\,s$^{-1}$ & $\arcsec$ & km\,s$^{-1}$ &  \\
			\hline                        
			\rule{0pt}{3ex}		
			
			H$\alpha$ 	&	$-$74\,$\pm$\,1	&	$-$161\,$\pm$\,2	& +36\,$\pm$\,2  & $-$112$\pm$\,1 &	$-$\,216$\pm$\,2  & $-$43$\pm$\,2 & 3.67 $\pm$ 0.04 & $-$93\,$\pm$\,1 & $\sim$0.44 \\		
			
			[N\,{\sc ii}]	& $-$75\,$\pm$\,1 &	$-$165\,$\pm$\,2	& +27\,$\pm$\,2 & $-$112$\pm$\,1 &	$-$\,212\,$\pm$\,2	& $-$49\,$\pm$\,2  & 3.98 $\pm$ 0.04 & $-$94\,$\pm$\,1 & $\sim$0.49 \\
			
			[S\,{\sc ii}]	& $-$73$\pm$\,2 &	$-$157\,$\pm$\,2	& +13\,$\pm$\,4 & $-$107$\pm$\,2 & $-$222\,$\pm$\,6	& $-$46\,$\pm$\,2 & 3.71 $\pm$ 0.11 & $-$90\,$\pm$\,1 & $\sim$0.82 \\
			
			[O\,{\sc iii}]	& $-$74$\pm$\,2 &	$-$120\,$\pm$\,2	& +28\,$\pm$\,3 & $-$113$\pm$\,2 & $-$200\,$\pm$\,13 & $-$54\,$\pm$\,11 & 3.79 $\pm$ 0.12 & $-$93\,$\pm$\,1 & $\sim$1.19 \\

			\hline                                  
		\end{tabular}
		\tablefoot{
			\tablefootmark{a}{North-western lobe (NW lobe) and south-eastern Lobe (SE lobe) as indicated in Figure \ref{fig:pv_map}. Velocities are reported with respect to the LSR.}
			\tablefootmark{b}{Velocity at the emission peak.}
			\tablefootmark{c}{Minimum ($V_{\rm min}$) and maximum ($V_{\rm max}$) radial velocity at 10$\%$\,$\pm$\,1$\sigma$ of the peak, where $\sigma$ is the noise of each PV map of Fig.\,\ref{fig:pv_map}.}
			\tablefootmark{d}{Relative distance between the emission peaks of the two bow-shocks.}
			\tablefootmark{e}{Systemic velocity, which is the mean value between the velocities at the emission peaks.}
            \tablefootmark{f}{Relative intensity between the emission of the NW lobe ($I_{\rm NW}$) and SE lobe ($I_{\rm SE}$).}
		}
	\end{table*}
In Table\,\ref{table:bow_shocks} we list the spectral, spatial, and emission properties of NW and SE lobes derived from the PV maps (Fig.\,\ref{fig:pv_map}). We measured the radial velocity of the emission peaks ($V_{\rm peak}$), the observed minimum ($V_{\rm min}$) and maximum ($V_{\rm max}$) radial velocities in the integrated spectra (see above) at the 10\% level of the intensity peak (FW0.1) (which avoids the insertion of noise in the [S\,{\sc ii}] and [O\,{\sc iii}] emission lines, and allows us to obtain a coherent value from the four emission lines), the angular separation between the emission peaks of the two lobes, as obtained from gaussian line fits to the spatial intensity distribution at the position of the intensity peaks, and the relative intensity of the emission peaks. Throughout this paper radial velocities will be quoted with respect to the Local Standard of Rest (LSR).

The emission peaks are separated by $\simeq$37 km\,s$^{-1}$, the NW lobe being redshifted with respect to the SE one (Table\,\ref{table:bow_shocks}). The centroid radial velocity of the two emission peaks is very similar in the four emission lines and their values are included in Table\,\ref{table:bow_shocks}; their mean value is $-$93$\pm$1 km\,s$^{-1}$ that will be considered the systemic velocity of the optical nebula. It is similar to, albeit somewhat larger than, the value of $-$85$\pm$2\,km\,s$^{-1}$ obtained by \cite{san12} from the CO emission, and the value of $-$80\,km\,s$^{-1}$ obtained by GL+91, perhaps due to the different spectral resolution. Secondary peaks are observed in the emission features, although more clearly in the SW lobe than in the NW one.

The FW0.1 in H$\alpha$ and [N\,{\sc ii}], amounts about 200 km\,s$^{-1}$ in NW lobe and about 170 km\,s$^{-1}$ in SE lobe. In [S\,{\sc ii}] and [O\,{\sc iii}], the values of FW0.1 are similar in both lobes and amount $\simeq$173 and $\simeq$147 km\,s$^{-1}$, respectively. It should be noted that the observed differences in FW0.1 are probably due to the different S/N ratio in the four emission lines.

The angular separation between the intensity peaks (Table\,\ref{table:bow_shocks}) is different in each emission line. However, given the errors in the [S\,{\sc ii}] and [O\,{\sc iii}] emission lines, and that the line profiles of H$\alpha$ and [N\,{\sc ii}] are very well defined, the only conclusion is that the angular separation between H$\alpha$ and [N\,{\sc ii}] presents differences, while the separation in [S\,{\sc ii}] and [O\,{\sc iii}] seems to be similar to that of H$\alpha$. Finally, the SE lobe is stronger than the NW one in H$\alpha$ and [N\,{\sc ii}], the contrary is observed in [O\,{\sc iii}], and both lobes present a similar intensity in [S\,{\sc ii}] (Table\,\ref{table:bow_shocks}).

	\section{Discussion} 

\subsection{The shock-excited emission from IRAS22568}
	\label{sec:bow_shocks}

As already mentioned, the electron temperatures obtained in 2021 are incompatible with photoionisation and suggest shock excitation. In order to understand the origin of the line emission, we have made use of the Mexican Million Models Database \citep[3MdB;][]{mor15}, which allows us to analyse whether the line intensities are consistent with shocks or photoionisation. In the Appendix\,\ref{ap:models} we explain the setup for the shock and photoionisation models employed below. 

In Fig.\,\ref{fig:i22568_shockmodels} we show the location of shock and photoionisation models in a $T_{\rm e}$-- $T_{\rm e}$ diagram (namely, an [O\,{\sc iii}]4363/5007 vs. [N\,{\sc ii}]5755/6584 diagram) and the location of IRAS22568 in 2021 (red star). We do not try to look for the exact models that reproduce the observed line ratio, we rather use the maximum number of models to define the whole region where they are located in the $T_{\rm e}$ (O\,{\sc iii})--$T_{\rm e}$(N\,{\sc ii}) diagram, relative to the type of ionisation. While the [N\,{\sc ii}]5755/6584 line intensity ratio can easily be reproduced by photoionisation and shock models, the high value of the [O\,{\sc iii}]4363/5007 line intensity ratio can only be reproduced by shock models, the highest ratio obtained by photoionisation models still being 0.25 dex below the observed value.

\textbf{\begin{figure}
		\centering
        \includegraphics[bb=11 30 383 670,width=6.6cm,clip]{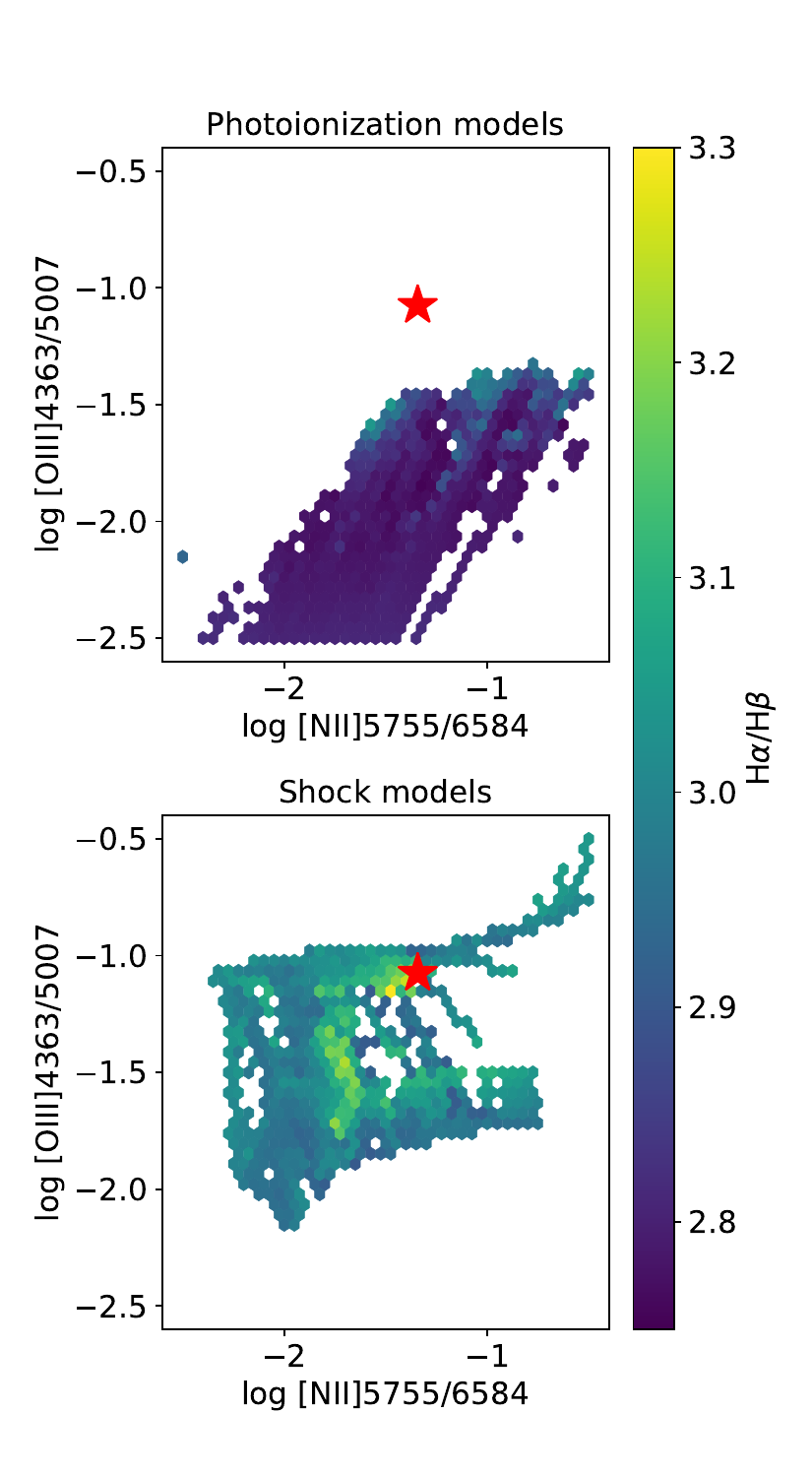}
		\caption{Location of models from the 3MdB database (see text for details) in an [O\,{\sc iii}]4363/5007 vs. [N\,{\sc ii}]5755/6584 diagram. Top panel shows models of PNe, bottom panel shows shock and shock+precursor models. The color code for the models corresponds to the H$\alpha$/H$\beta$ line intensity ratio. The red star is the location of the 2021 observations of IRAS22568.}
		\label{fig:i22568_shockmodels}
	\end{figure}}
 
If shocks were present in 2021, $T_{\rm e}$ cannot be determined in the classical manner and the values listed in Table\,1 should be seen with caution. Determining $T_{\rm e}$ requires detailed models involving shocks \citep[see e.g.,][]{mon22}, which are deferred to a future paper.  We also noticed that if shock excitation exists, the H$\alpha$/H$\beta$ theoretical ratio to determine $c$(H$\beta$) could be higher than the assumed value of 2.85. Thus, we have calculated $c$(H$\beta$) and intrinsic emission line intensities for an H$\alpha$/H$\beta$ ratio of 3.1, but we obtained values that are very similar to and within the errors of those in Table\,\ref{table:em_lines}.

The shocks observed in 2021 most probably are the mechanism responsible for the large velocity widths observed in the PV maps from 2023 (Fig.\,\ref{fig:pv_map}). In fact, emission lines with large velocity widths are well known in Herbig-Haro objects associated with young stars \citep[e.g.][]{strom74, dop78, raga86, har87}. Nowadays it is well established that this kind of emission features trace bow-shocks associated with high-velocity jets or bullets and their interaction with the surrounding medium. 

    Assuming the radiative bow-shock models by \cite{har87}, the shock velocity and viewing angle of NW lobe and SE lobe (Fig.\,\ref{fig:pv_map}) can be determined in a very simple way. We will concentrate in the H$\alpha$ and [N\,{\sc ii}] data that are of better quality, and note that the [S\,{\sc ii}], and [O\,{\sc iii}] data provide compatible results. The shock velocity is the difference between the observed $V_{\rm min}$ and $V_{\rm max}$ in Table\,3. In the NW lobe the bow-shock velocity and viewing angle with respect to the observer are $\simeq$195\,km\,s$^{-1}$ and $\simeq$106$\pm$2$\degr$ with respect to the line of sight, respectively; in the SE lobe the corresponding values are $\simeq$170\,km\,s$^{-1}$ and $\simeq$64$\pm$3$\degr$. Using the radial velocity of the intensity peaks (with respect to the systemic velocity) as representative of the velocity of the bow-shocks, we obtained velocities of $\simeq$70 and $\simeq$45\,km\,s$^{-1}$ in the NW and SE lobe, respectively.

    We tried to identify which nebular features in the \textit{HST} image (Fig.\,\ref{fig:i22568_hst_caha}) could be associated with the bow-shocks, but IRAS22568 shows many knots and curved filaments, and, given the small angular size of the lobes, comparable to the spatial resolution of the MES spectra, it is difficult to pinpoint an association. On the other hand, we cannot rule out that the spectral features are an overlapping of several bow-shocks associated with different knots and filaments in the bipolar lobes. If so, the viewing angle and expansion velocities obtained above may represent averaged values for each lobe, weighted by the contribution of each individual bow-shock. This may explain the differences in expansion velocity and tilt angle of each lobe without implying that the lobes have a different orientation.
    
    A comparison of the H$\alpha$ and [N\,{\sc ii}] PV maps shown in Fig.\,\ref{fig:pv_map} with those by GL+91 (their Fig.\,4) is revealing. Despite the differences in spatial and velocity resolution between both sets of PV maps, those by GL+91 show the emission peaks of each lobe, a secondary peak in the SE lobe, and a velocity width of $\simeq$350--380 km\,s$^{-1}$ at the spatial position of the intensity peaks, as measured at the weakest contour level. These results agree very well with ours, suggesting that bow-shock excitation could already be present in 1988. Moreover, our high-resolution spectra include a recombination (H$\alpha$), a high-excitation ([O\,{\sc iii}]), and two low-excitation ([N\,{\sc ii}] and [S\,{\sc ii}]) emission lines and, in all four cases, the emission features are compatible with bow-shock excitation. Therefore, it is possible that the optical spectrum of IRAS22568 in the three epochs is dominated by shocks. This raises doubts on the classification of IRAS22568 as a low-ionisation PN. Nevertheless, concluding that IRAS22568 is a post-AGB nebula could be premature because photoionised gas could be masked by shocks that are the dominant excitation mechanism. Furthermore, as an additional result, the existence of shocks might be intimately related with the non-thermal radio continuum emission from IRAS22568. Finally, it is worth to emphasize that, besides the $T$$_{\rm e}$-sensitive line ratios in 2021 (Fig.\,\ref{fig:i22568_shockmodels}), high resolution spatially resolved spectroscopy has been crucial to determine the excitation mechanism in 2023: without this kind of data, post-AGB nebula with shock-excited emission lines could be confused with low-ionisation young PNe, when observed at low- or intermediate-spectral resolution. 

    IRAS22568 belongs to a small group of post-AGB nebulae and PNe that present emission features with large velocity widths (typically $\geq$150\,km\,s$^{-1}$) characteristics of bow-shock excitation. By means of high-resolution optical spectroscopy, large velocity widths have been detected in the PNe Hen\,2-111 \citep{mea89}, M\,1-16 \citep{schw92, gomu23}, KjPn\,8 \citep{lop95}, M\,2-48 \citep{vaz00, lopm02}, and IRAS\,18061$-$2505 \citep{mir21}. Other PNe present bow-shaped structures that have been successfully reproduced using bow-shock models, although in these cases the structures do not exhibit such large velocity widths (typically 40--90\,km\,s$^{-1}$) and they are irradiated by the central star; examples include K\,4-47, IC\,4634, and Hu\,1-2 \citep[see also Mari et al. 2023 for more similar PNe]{gonc04, mar08, mir12, fang15}. Similar emission features are observed in the post-AGB nebulae M\,1-92 \citep{solf94}, OH\,231.8+04.2 \citep{san00}, Hen\,3-1475 \citep{rie03}, M\,2-56 \citep{san10}. The reason for the scarcity of large velocity widths associated with jets in PNe in not clear, although it could be related to photoionisation being the dominant excitation mechanism, which (partially) hides the shock-excited emission, and/or to a weakness of the shock contribution with time.

 \begin{figure}
		\centering
		\begin{subfigure}{0.92\linewidth}
		\includegraphics[width=\linewidth]{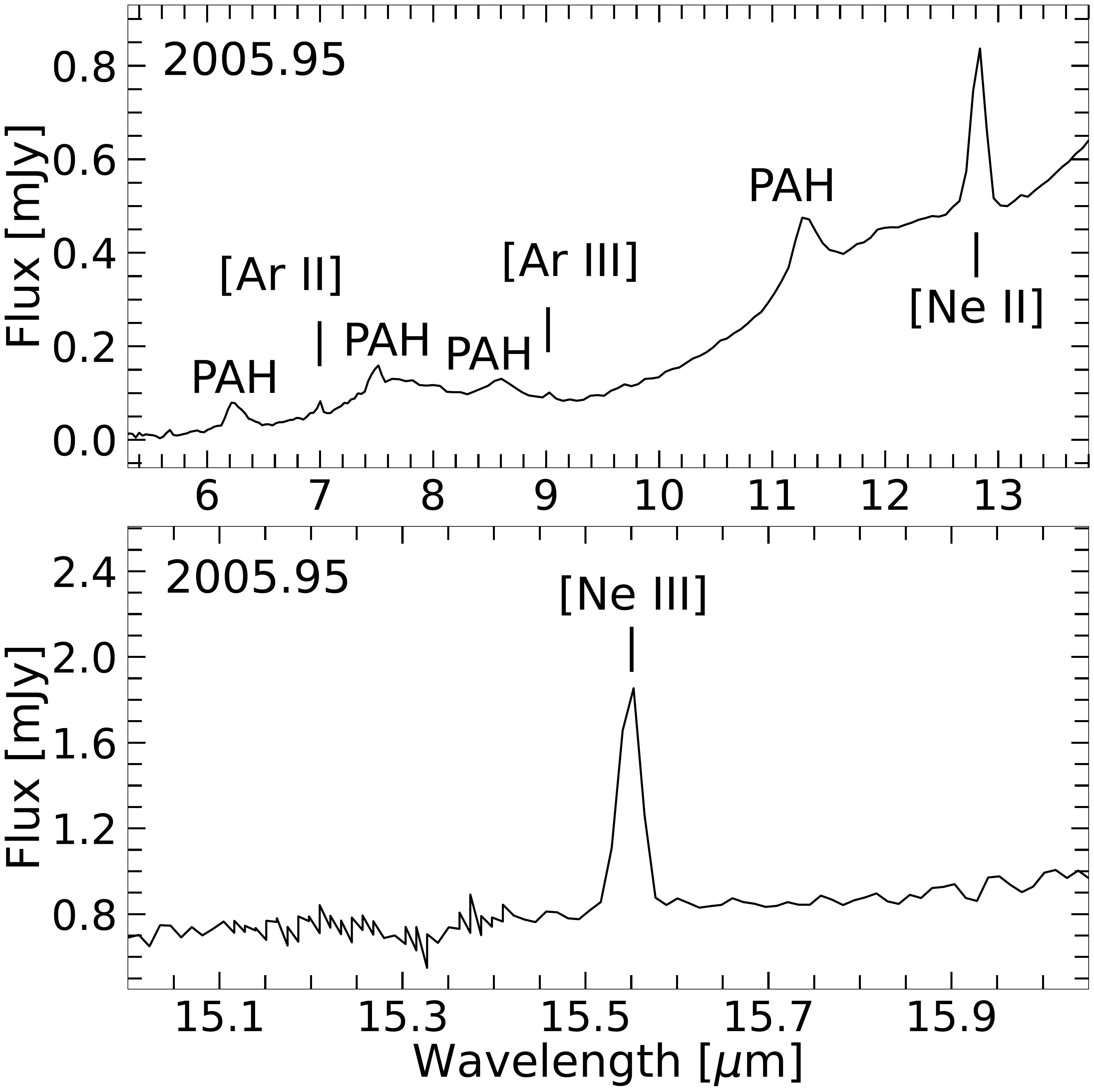}
		\end{subfigure}
		\caption{\textit{Spitzer} spectra of IRAS22568 obtained on 2005.95. Forbidden emission lines of ionised gas ([Ar\,{\sc ii}]$\lambda$7.0$\mu$m, [Ar\,{\sc iii}]$\lambda$9.0$\mu$m, [Ne\,{\sc ii}]$\lambda$12.8$\mu$m, and [Ne\,{\sc iii}]$\lambda$15.55$\mu$m) and broad emission of PAHs (6.4, 7.7, 8.6 and 11.3\,$\mu$m) are indicated.}
		\label{fig:spitzer}
\end{figure}

    \subsection{On the origin of the variability}

    The $r$ light curve (Fig.\,\ref{fig:i22568_r_photo}) shows that IRAS22568 underwent an energetic event around 1990, which increased its brightness. The consequences of that event are perceptible until the present. The radio continuum emission from the nebula has been fading since at least 2005 \citep{cer17} and, according to our spectra, IRAS22568 is recombining and cooling down between 2021 and 2023. Moreover, we consulted the \textit{Spitzer} Heritage Archive, and found spectra of IRAS22568 from 2005.95, which are shown in Fig.\,\ref{fig:spitzer}. These spectra display, among others, the [Ne\,{\sc iii}]15.6$\mu$m emission line which indicates that the nebular excitation was higher in 2006 than in later epochs (2021 and 2023) when the [O\,{\sc iii}] emission lines were relatively weak. Consequently, it is quite possible that the [O\,{\sc iii}] lines were stronger in 2006 than in 2021, and that IRAS22568 is recombining and cooling down since 2006 and even earlier. Finally, the spectral and kinematical properties provide clear evidence that the nebular emission is dominated by shocks in 2021 and 2023 and, perhaps, in 1988. 

   The variability of IRAS22568 could be due to changes of the physical conditions in the shocks. New ejections could have reached and interacted with the lobes by 1990. Similarly, already ejected material could encounter high density regions in the lobes and be decelerated by that date. In both cases, a rapid increase in $T_{\rm e}$ and $N_{\rm e}$ should be expected, resulting in an increase of the emission line fluxes and excitation. Once the interaction weakens or ceases, recombination and cooling down in the post-shock region would lead to a decrease in the excitation and brightness of the nebula, as observed since 2005-2006. Numerical simulations are necessary to check whether this scenario can account for the detailed observed variability and the involved time scales, and whether the physical conditions required for it to occur are reasonable for PNe. We note that, if the variability is due to changes in the shocks only, IRAS22568 would be a post-AGB nebula or proto-PN similar to, for example, M\,1-92, rather than a PN.

   Alternatively, a nova-like eruption could have been responsible for the sudden increase of brightness of IRAS22568. This kind of event is typical of binary systems hosting an AGB star and an accreting white dwarf. Such symbiotic binaries are in a different evolutionary phase than post-AGB stars and PNe. To check whether IRAS22568 could be related to this kind of binaries, we have built its spectral energy distribution (SED; see Appendix\,\ref{sec:sed}) that is shown in Fig.\,\ref{fig:sed}. The SED presents the emission peak at $\lambda$\,$\simeq$\,30\,$\mu$m, which is typical of optically obscured post-AGB stars and young PNe \citep[e.g.][]{ram09, ram12}, but different from that of symbiotic binaries, where the peak of the SED usually lies at $\lambda$\,$\simeq$1\,$\mu$m \citep[e.g.][]{rod14}. Although more observations would be necessary, the SED of IRAS22568 does not seem to favour a nova scenario as the cause of its variability.

    Finally, a third possibility is a late-thermal pulse (LTP) occurred around 1990, as it is suggested by the remarkable similarities between the properties and temporal evolution of IRAS22568 and those of the Stingray nebula (Hen\,3-1357) whose central star, SAO\,244567, has been proposed to undergo an LTP \citep{rein14, rein17}. The light curves of IRAS22568 (Fig.\,\ref{fig:i22568_r_photo}) and Hen\,3-1357 \citep{sch15} are similar to each other. In both objects, the spectral variability of the optical emission lines is consistent with cooling down and recombination during decades \citep[see Sec.\,\ref{section:physical} and][]{ark13, pena22, bal21}. Their radio continuum emission have been characterised by a fading of the flux density in the last decades, and negative spectral indices indicating non-thermal emission \citep[][]{cer17, harv18}. As already mentioned, non-thermal emission and shocks appear related to each other in IRAS22568 (Sect\,\ref{sec:bow_shocks}). The same may hold for Hen\,3-1357 where shocks have been suggested to occur along the nebular perimeter \citep{bal21}. Similarly, a relationship between shocks and non-thermal radio continuum emission is observed after the (very late) thermal pulse occurred in Sakurai's object \citep{haj24}. Moreover, if an LTP in IRAS22568 has generated a shock wave that propagates into the previous material, all the knots and filaments would be shocked, increasing the nebular brightness, excitation and the subsequent recombination and cooling down. Nevertheless, in the LTP scenario, the situation may be more complicated. The increase of the effective temperature of the central star after the LTP \citep[and references therein]{law23} would provide an additional source of ionising photons and the emission line spectra of IRAS22568 would be produced in a combination of shocks and photoionisation, each presenting its own timescale of variability depending on the evolution of the central star and nebula. It is interesting to note that the \textit{Spitzer} spectra (Fig.\,\ref{fig:spitzer}) display broad emission features attributed to dust made of polycyclic aromatic hydrocarbons (PAHs), which might be fresh byproducts of the LTP, and indicate carbon-rich material in the nebula. Finally, if the LTP scenario is correct, IRAS22568 would indeed be a nascent PNe.

    In contrast with SAO\,244567 that is visible and its properties and evolution have been studied for decades \citep{hen76,parth93, parth95, rein14, rein17}, the central star of IRAS22568 has never been observed and its current evolutionary stage is unknown. In particular, within the LTP scenario, we cannot assert whether the central star of IRAS22568 is increasing its effective temperature or returning back to the AGB \citep[see][]{haj20}. Millimeter observations could elucidate whether the molecular abundances of IRAS22568 are similar to those in other sources in which an LTP seems to have occurred \citep[e.g.,][]{sch18}. In addition, a photometric and spectroscopic monitoring in the coming years and modelling of the data are key to trace the evolution of IRAS22568 and confirm the origin of the variability. 
    
    \section{Conclusions}

  IRAS22568 has been classified as a low-ionisation PN and shows non-thermal radio continuum emission, suggesting to be a nascent PN. We performed on this source intermediate-resolution optical spectroscopy in 2021 and 2023, high-resolution spectroscopy in 2023, and gathered a discrete \textit{r}-filter light curve from 1953 to 2020. 
  
  It is clear that an energetic event occurred in IRAS22568 around 1990, which manifested itself as a sudden and rapid increase of the nebular brightness. Afterwards, the brightness has been diminishing much slower. As compared with a published spectrum from 1988, ours of 2021 shows a noticeable increase of the emission line fluxes and the presence of the [O\,{\sc iii}] emission lines which were not detected in 1988. Two years later, in 2023, the emission line fluxes decreased and the [O\,{\sc iii}] emission lines almost vanished. Actually, the observed spectral variability between 2021 and 2023 shows that IRAS22568 is recombining and cooling down, a phenomenon that is probably occurring since at least 2005, as indicated by the reported fading of the radio continuum emission and archival mid-IR spectra around that epoch. 
  
  Some emission line ratios in 2021 reveal that the spectrum is shock-excited. Moreover, the large velocity widths ($\sim$400\,km\,s$^{-1}$ ) observed in the high-resolution 2023 spectra are compatible with bow-shock excitation, which could also be present in 1988. Our data suggest a connection between shock excitation and non-thermal radio continuum emission. Nevertheless, the existence of some contribution of photoionisation to the spectra cannot be discarded at least since 2006. 

  Changes of the physical conditions in the shocks could eventually explain the observed variability but modelling is necessary to confirm this scenario. A nova-like eruption would be another possibility, although this is not favored by the SED of IRAS22568. Finally, a late thermal pulse in the central star of IRAS22568 provides a simple and consistent explanation for the energetic event in 1990 and the subsequent photometric and spectral evolution of the object. In fact, we found noticeable similarities between the variability of IRAS22568 and that of the Stingray nebula, which has been interpreted as due to a late thermal pulse.

	\begin{acknowledgements} We are very grateful to our anonymous referee for his/her constructive comments that have improved the paper. We thank Calar Alto Observatory for allocation of director's discretionary time to this programme. This research is based on observations collected at the Centro Astron{\'o}mico Hispano en Andaluc{\'i}a (CAHA) at Calar Alto, operated jointly by Junta de Andaluc{\'i}a and Consejo Superior de Investigaciones Cient{\'i}ficas (IAA-CSIC); upon observations at the Observatorio Astron\'omico Nacional on the Sierra San Pedro M\'artir (OAN-SPM), Baja Califormia, M\'exico; on data from the program 9463 of the NASA/ESA Hubble Space Telescope obtained from the Space Telescope Science Institute, which is operated by the Association of Universities for Research in Astronomy, Inc., under NASA contract NAS 5-26555; on data from the VLA and the National Radio Astronomy Observatory is a facility of the National Science Foundation operated under cooperative agreement by Associated Universities, Inc. IRAF used here is distributed by the National Optical Astronomy Observatories, which is operated by the Association of Universities for Research in Astronomy, Inc., under contract to the National Science Foundation. The POSS-II were produced at the Space Telescope Science Institute under U.S. Government grant NAG W-2166. The images of these surveys are based on photographic data obtained using the Oschin Schmidt Telescope on Palomar Mountain and the UK Schmidt Telescope. The plates were processed into the present compressed digital form with the permission of these institutions.The Guide Star Catalogue-II is a joint project of the Space Telescope Science Institute and the Osservatorio Astronomico di Torino. Space Telescope Science Institute is operated by the Association of Universities for Research in Astronomy, for the National Aeronautics and Space Administration under contract NAS5-26555. The participation of the Osservatorio Astronomico di Torino is supported by the Italian Council for Research in Astronomy. Additional support is provided by European Southern Observatory, Space Telescope European Coordinating Facility, the International GEMINI project and the European Space Agency Astrophysics Division. The PS1 and the PS1 public science archive have been made possible through contributions by the Institute for Astronomy, the University of Hawaii, the Pan-STARRS Project Office, the Max-Planck Society and its participating institutes, the Max Planck Institute for Astronomy, Heidelberg and the Max Planck Institute for Extraterrestrial Physics, Garching, The Johns Hopkins University, Durham University, the University of Edinburgh, the Queen's University Belfast, the Harvard-Smithsonian Center for Astrophysics, the Las Cumbres Observatory Global Telescope Network Incorporated, the National Central University of Taiwan, the Space Telescope Science Institute, the National Aeronautics and Space Administration under Grant No. NNX08AR22G issued through the Planetary Science Division of the NASA Science Mission Directorate, the National Science Foundation Grant No. AST-1238877, the University of Maryland, Eotvos Lorand University (ELTE), the Los Alamos National Laboratory, and the Gordon and Betty Moore Foundation. Based on observations obtained with the Samuel Oschin 48-inch Telescope at the Palomar Observatory as part of the ZTF project. ZTF is supported by the National Science Foundation under Grant No. AST-1440341 and a collaboration including Caltech, IPAC, the Weizmann Institute for Science, the Oskar Klein Center at Stockholm University, the University of Maryland, the University of Washington, Deutsches Elektronen-Synchrotron and Humboldt University, Los Alamos National Laboratories, the TANGO Consortium of Taiwan, the University of Wisconsin at Milwaukee, and Lawrence Berkeley National Laboratories. Operations are conducted by COO, IPAC, and UW. This work is based [in part] on archival data (ID: 20590, PI: R.\,Sahai) obtained with the Spitzer Space Telescope, which was operated by the Jet Propulsion Laboratory, California Institute of Technology under a contract with NASA. Support for this work was provided by NASA. This work has made use of the SIMBAD database, operated at the CDS, Strasbourg, France, and the NASA/IPAC Infrared Science Archive, which is operated by the Jet Propulsion Laboratory, California Institute of Technology, under contract with the National Aeronautics and Space Administration. It also makes use of data products from 2MASS (a joint project of the University of Massachusetts and the Infrared Processing and Analysis Center/California Institute of Technology, funded by NASA and the NSF), AKARI (a JAXA project with the participation of ESA), HERSCHEL (Herschel is an ESA space observatory with science instruments provided by European-led Principal Investigator consortia and with important participation from NASA), IRAS (was a joint project of the US, UK and the Netherlands), MSX (funded by the Ballistic Missile Defense Organization with additional support from NASA Office of Space Science), and WISE (a joint project of the University of California, Los Angeles, and the Jet Propulsion Laboratory/California Institute of Technology, funded by the NASA). RAC, LFM, JFG are supported by grants PID2020-114461GB-I00, PID2023-146295NB-I00, and CEX2021-001131-S, funded by MCIN/AEI /10.13039/501100011033, and by grant P20-00880, funded by the Economic Transformation, Industry, Knowledge and Universities Council of the Regional Government of Andalusia and the European Regional Development Fund from the European Union. RC also acknowledges support by the predoctoral grant PRE2018-085518, funded by MCIN/AEI/ 10.13039/501100011033 and by ESF Investing in your Future. CM acknowledges the support of UNAM/DGAPA/PAPIIT grants IN101220 and IG101223.
	\end{acknowledgements}
	
	%
	%

\begin{appendix}

\section{Atomic data sets used in PyNeb} 

Table\,\ref{tab:atomic_data} shows the atomic data sets used for collisionally-excited lines in the PyNeb calculations presented in this paper.

\begin{table}
\caption{Atomic data sets used for collisionally-excited lines in the PyNeb calculations. \label{tab:atomic_data}}
\begin{tabular}{lcc}
\hline
Ion & Transition Probabilities\tablefootmark{a} & Collision Strengths\tablefootmark{b} \\
\hline
N$^{+}$   &  1 & 1 \\
O$^{2+}$  &  1, 2 & 2, 3 \\
S$^{+}$   &  3 & 3 \\
\hline
\end{tabular}

\tablefoot{\tablefootmark{a}{1. \citet{fro04}. 2. \citet{sto00} 3. \citet{ryn19}.}
\tablefootmark{b}{1. \citet{tay11}.  2. \citet{fro04}. 3. \citet{sto14}}. 4. \citet{tay10}.}

\end{table}

\section{Emission line fluxes of IRAS22568 in the three epochs.}

In Table\,\ref{tab:lines_uncorrected} we report the emission line fluxes in 1988, 2021, and 2023. The emission line fluxes in 1988 are taken from GL+91 (their Table\,2, column 4).

\begin{table}[b]
\caption{Absolute emission line fluxes F($\lambda$) in units of 10$^{-15}$ erg\,cm$^{-2}$\,s$^{-1}$ on IRAS22568 in the three epochs.} \label{tab:lines_uncorrected}
\centering
\begin{tabular}{lccc}
\hline \hline
    & 1988.58 & 2021.58 & 2023.88 \\
Emission line   &  F($\lambda$)\tablefootmark{a}  &  F($\lambda$)\tablefootmark{a}  & F($\lambda$)\tablefootmark{a} \\

\hline
4101	H$\delta$	&	--			&	2.80$\pm$0.14	&	--			\\
4340	H$\gamma$	&	--			&	3.93$\pm$0.24	&	--			\\
4363	[O\,{\sc iii}]	&	--			&	0.61$\pm$0.13	&	$<$34.4		\\
4471	He\,{\sc i}	&	--			&	0.52$\pm$0.04	&	--			\\
4861	H$\beta$	&	1.16$\pm$0.11&	6.88$\pm$0.11	&	4.11$\pm$0.11	\\
4959	[O\,{\sc iii}]	&	--			&	2.80$\pm$0.08	&	0.62$\pm$0.06	\\
5007	[O\,{\sc iii}]	&	--			&	8.19$\pm$0.09	&	1.67$\pm$0.08	\\
5755	[N\,{\sc ii}]	&	0.48$\pm$0.04	&	0.757$\pm$0.019	&	0.314$\pm$0.012	\\
5876	He\,{\sc i}]	&	0.86$\pm$0.07	&	0.952$\pm$0.017	&	0.529$\pm$0.016	\\
6300	[O\,{\sc i}]	&	3.41$\pm$0.17	&	1.531$\pm$0.013	&	2.760$\pm$0.028	\\
6363	[O\,{\sc i}]	&	1.43$\pm$0.07	&	0.557$\pm$0.012	&	1.105$\pm$0.026	\\
6548	[N\,{\sc ii}]	&	13.22$\pm$0.23	&	4.82$\pm$0.16	&	3.30$\pm$0.07	\\
6563	H$\alpha$	&	51.9$\pm$0.9	&	19.61$\pm$0.17	&	11.71$\pm$0.07	\\
6584	[N\,{\sc ii}]	&	40.3$\pm$0.9	&	15.14$\pm$0.16	&	9.66$\pm$0.07	\\
6678	He\,{\sc i}	&	0.58$\pm$0.07	&	0.254$\pm$0.007	&	0.176$\pm$0.010	\\
6716	[S\,{\sc ii}]	&	1.91$\pm$0.16	&	0.482$\pm$0.008	&	0.537$\pm$0.011	\\
6731	[S\,{\sc ii}]	&	3.84$\pm$0.27	&	0.925$\pm$0.009	&	1.069$\pm$0.011	\\
7002	O\,{\sc ii}	&	--			&	0.034$\pm$0.0011	&	--			\\
7065	He\,{\sc i}	&	1.28$\pm$0.11	&	0.319$\pm$0.005	&	0.193$\pm$0.008	\\
7137	[Ar\,{\sc iii}]	&	1.97$\pm$0.25	&	0.438$\pm$0.005	&	0.131$\pm$0.007	\\
7150	[Fe\,{\sc ii}]	&	--			&	0.103$\pm$0.005	&	0.173$\pm$0.007	\\
7230	[Fe\,{\sc ii}]	&	--			&	0.041$\pm$0.003	&	0.145$\pm$0.010	\\
7254	O\,{\sc i}	&	--			&	0.037$\pm$0.003	&	0.133$\pm$0.010	\\
7281	He\,{\sc i}	&	--			&	0.052$\pm$0.006	&	0.242$\pm$0.008	\\
7319	[O\,{\sc ii}]	&	blended\tablefootmark{b}			&	1.375$\pm$0.010	&	0.623$\pm$0.009	\\
7330	[O\,{\sc ii}]	&	blended\tablefootmark{b}	&	1.057$\pm$0.010	&	0.573$\pm$0.009	\\
7378	[Ni\,{\sc ii}]	&	0.70$\pm$0.14	&	0.090$\pm$0.009	&	0.125$\pm$0.006	\\
7447	[Fe\,{\sc ii}]	&	--			&	0.0383$\pm$0.0021	&	0.074$\pm$0.007	\\
7751	[Ar\,{\sc iii}]	&	--			&	0.126$\pm$0.005	&	0.0280$\pm$0.0020	\\
8446	O\,{\sc i}	&	blended\tablefootmark{c} &	0.229$\pm$0.004	&	0.066$\pm$0.004	\\
8467	P17	&	blended\tablefootmark{c} &	0.056$\pm$0.003	&0.0294$\pm$0.0021	\\
8502	P16	&	--			&	0.058$\pm$0.004	&0.063$\pm$0.004	\\
8545	P15	&	--			&	0.075$\pm$0.004	&	0.081$\pm$0.005	\\
8578	[Cl\,{\sc ii}]	&	--			&	0.053$\pm$0.004	&	0.093$\pm$0.005	\\
8598	P14	&	--			&	0.054$\pm$0.004	&	0.072$\pm$0.006	\\
8605	[Fe\,{\sc ii}]	&	--			&	0.090$\pm$0.004	&	0.180$\pm$0.005	\\
8665	P13	&	--			&	0.087$\pm$0.004	&	0.093$\pm$0.005	\\
8750	P12	&	--			&	0.066$\pm$0.003	&	0.094$\pm$0.005	\\
8863	P11	&	--			&	0.105$\pm$0.003	&	0.111$\pm$0.004	\\
8890	[Fe\,{\sc ii}]	&	--			&	0.063$\pm$0.003	&	0.094$\pm$0.004	\\
9015	P10	&	3.0$\pm$0.4	&	0.126$\pm$0.005	&	0.067$\pm$0.003	\\
9069	[S\,{\sc iii}]	&	19.1$\pm$1.6	&	0.992$\pm$0.005	&	0.212$\pm$0.004	\\
9229	P9	&	7.4$\pm$0.8	&	0.237$\pm$0.005	&	0.210$\pm$0.006	\\
9535	[S\,{\sc iii}]	&	51.04$\pm$3.26\tablefootmark{d}	&	2.217$\pm$0.013	&	0.487$\pm$0.012	\\

\hline

\hline                                   
\end{tabular}
\tablefoot{\tablefootmark{a}{'--' stands for not detected.}
\tablefootmark{b}{[O\,{\sc ii}]\,$\lambda$7319,7330 are blended with $F$(sum)=12.6$\pm$0.8.}
			\tablefootmark{c}{O\,{\sc i}\,$\lambda$8446 and P\,17\,$\lambda8467$ are blended with $F$(sum)=5.8$\pm$0.5.}
            \tablefootmark{d}{Blended with P\,8\,$\lambda9546$.}
                        }	
		
\end{table}

\section{Parameters of the shock and photoionisation models in the 3MdB}
\label{ap:models}

\subsection{Shock models}

The shock models are taken from the work of \citet{ala19} who computed models using Mappings V \citep{sut17} and store them in the 3MdB \citep{ala19}. These models use the same grid definition as \citet{all08}. We consider here the 15968 models that correspond to the "shock" and the "shock+precursor" results from the "Allen08" shock family, the precursor being the region photoionised by the photons emitted by the shock itself, for velocities higher than 100 km\,s$^{-1}$ (at lower velocities no ionising photons are produced by the shock). This preionisation is treated by Mappings in a fully consistent way and is described in detail by \citet{sut17}. The free parameters of the grid are the metallicities, the pre-shock density, the shock velocity and the magnetic field. The details of the parameter distribution are given in \citet{ala19}.

\subsection{Photoionisation models}

The PN photoionisation models have been computed by \citet{del14} using the multi-purpose transfer code CLOUDY v17 \citep{fer17}. These models are taken from the 3MdB database, using the reference \verb|ref LIKE "PNe_202_"|. We filter the models using the \verb|com6=1 AND MassFrac > 0.7| commands. The first one is to only consider the realistic PNe models as defined in Sec. 2.1 of \citet{del14} and in the grid model webpage\footnote{https://sites.google.com/site/mexicanmillionmodels}. The second filter is used to only select the radiation-bonded models, as well as the matter-bounded models in which the mass of the nebula is greater than 70\% of the corresponding radiation-bounded case. The input parameters of the grid are: the nebular density and metallicity, the luminosity and effective temperature of the ionising spectral energy distribution (which can be a Planck function or a stellar atmosphere model), the presence or absence of dust, the distance between the central ionising source and the inner radius of the nebula. This leads to 68041 models.

\section{Spectral energy distribution (SED) of IRAS22568} 
\label{sec:sed}

\begin{figure}
		\centering
		\begin{subfigure}{0.93\linewidth}
		\includegraphics[width=\linewidth]{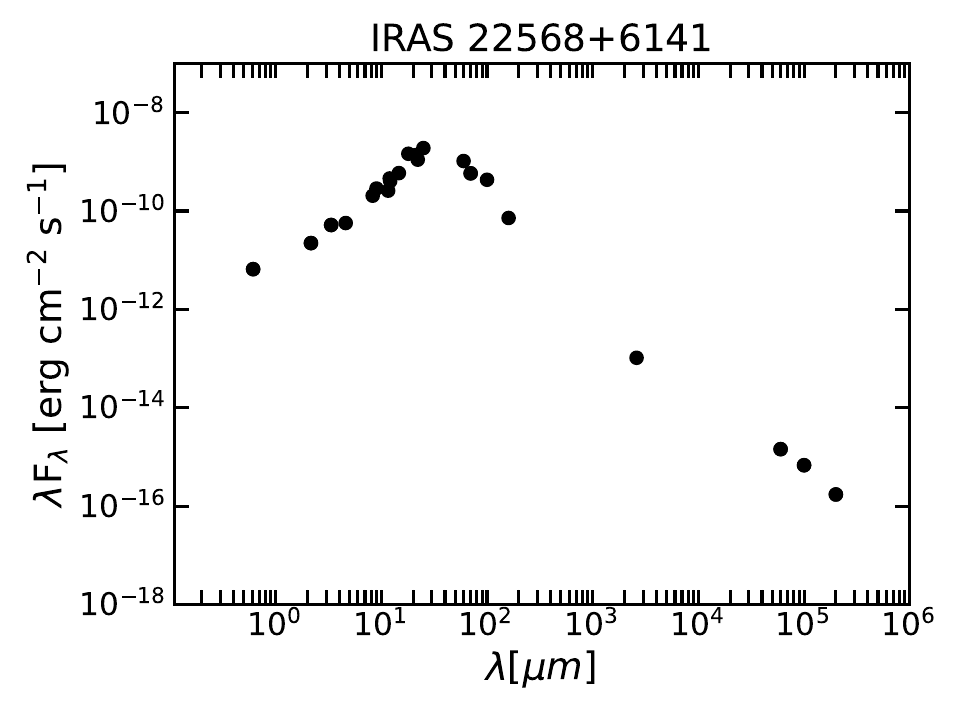}
		\end{subfigure}
		\caption{SED of IRAS22568. The size of the circles is larger than the errors.}
		\label{fig:sed}
\end{figure}

We have gathered available archival optical, infrared, millimeter, and centimeter data ($\sim$0.5–2$\times$10$^{5}$\,$\mu$m), and have built the SED of IRAS22568 (Fig.\,\ref{fig:sed}). Infrared data were obtained from the Two Micron All Sky Survey (2MASS), AKARI, \textit{Herschel} Space Observatory, Infrared Astronomical Satellite (IRAS), Midcourse Space Experiment (\textit{MSX}), and Wide-field Infrared Survey Explorer (all-\textit{WISE} survey). The flux density at 0.1 and 2.6\,mm were taken from \cite{mar24} and \cite{san12}, respectively. The flux density of the optical \textit{r}-magnitude corresponds to the average value of those presented in the Table\,\ref{table:photo}. Similarly, the radio flux density values are the average values reported by \cite{cer17} at 4.8 and 8.6 GHz, while the value at 1.4 GHz was reported by \cite{con98}. 

\end{appendix}
 
\end{document}